\documentclass[lettersize,journal]{IEEEtran}
\usepackage{amsmath,amsfonts}
\usepackage{algorithmic}
\usepackage{algorithm}
\usepackage{array}
\usepackage{siunitx}
\usepackage[caption=false,font=normalsize,labelfont=sf,textfont=sf]{subfig}
\usepackage{textcomp}
\usepackage{stfloats}
\usepackage{url}
\usepackage{chemformula}
\usepackage{verbatim}
\usepackage{graphicx, xcolor}
\usepackage{cite}

\begin{document}

\title{Synchronization in pairs of opto-thermally driven mechanically coupled micro-oscillators}

\author{Aditya Bhaskar, Mark Walth, Richard H. Rand, Alan T. Zehnder

\thanks{This material is based upon work supported by the National Science Foundation under Grant No. CMMI-1634664. The experimental work was performed in part at the Cornell NanoScale Facility, a member of the National Nanotechnology Coordinated Infrastructure (NNCI), which is supported by the National Science Foundation (Grant NNCI-2025233). The numerical work used the Extreme Science and Engineering Discovery Environment (XSEDE), which is supported by National Science Foundation grant number ACI-1548562. Specifically, it used the Bridges-2 system, which is supported by NSF award number ACI-1928147, at the Pittsburgh Supercomputing Center (PSC). This work also made use of the Cornell Center for Materials Research Shared Facilities which are supported through the NSF MRSEC program (DMR-1719875).}

\thanks{Aditya Bhaskar and Alan T. Zehnder are with the Sibley School of Mechanical and Aerospace Engineering, Cornell University, Ithaca, NY 14853 USA (email: ab2823@cornell.edu; atz2@cornell.edu).}

\thanks{Mark Walth is with the Department of Mathematics, Cornell University, Ithaca, NY 14853 USA (email: msw283@cornell.edu).}

\thanks{Richard H. Rand is with the Sibley School of Mechanical and Aerospace Engineering and the Department of Mathematics, Cornell University, Ithaca, NY 14853 USA (email: rhr2@cornell.edu).}

}

\maketitle

\begin{abstract}

We study the phenomenon of synchronization in pairs of doubly clamped, mechanically coupled silicon micro-oscillators. A continuous-wave laser beam is used to drive the micro-beams into limit cycle oscillations and to detect the oscillations using interferometry. Devices of different dimensions are used to introduce frequency detuning, and short silicon bridges connecting the micro-beams are used as mechanical coupling between the oscillators. The region of synchronization is plotted for the MEMS system in the detuning vs. coupling parameter space and compared with the numerical analysis of a corresponding, lumped-parameter model. Three states of oscillations are observed i.e. the drift state, quasi-periodic state, and the synchronized state. The numerical model also distinguishes between in-phase and out-of-phase synchronization where out-of-phase synchronization is observed at low coupling strengths and low frequency detuning. 
   
\end{abstract}


\section{Introduction} \label{sec:Introduction}

\IEEEPARstart{S}{ynchronization} in coupled oscillators has captured the imagination of scientists and engineers from the time it was first observed by Christiaan Huygens as ``sympathy of clocks” \cite{pikovsky2001universal}. Since then, the phenomenon of two or more oscillators spontaneously synchronizing in the presence of a sufficiently strong coupling field has been recorded in various contexts. Examples of synchronization in biological oscillators include swarms of rhythmically flashing fireflies \cite{winfree1980geometry}, coordinated cilia in \textit{trichoplax adhaerens} to achieve locomotion \cite{bull2021excitable}, the behavior of intestinal muscles \cite{diamant1970computer}, and light-based communication in diatom colonies \cite{font2021pelagic}. Inspired by nature, micro- and nano-scale oscillating devices have been engineered to exploit the phenomenon of synchronization to reduce phase noise \cite{zhang2015synchronization}, perform mechanical computing \cite{hoppensteadt2001synchronization, csaba2020coupled}, and sense acceleration \cite{xu2020programmable}. MEMS (micro-electro-mechanical systems) and NEMS (nano-electro-mechanical systems) oscillators have also been used to study the dynamics of nonlinear systems. Due to their dimensions in the sub-micron range, such devices exhibit strong geometric nonlinearities caused by large deformations \cite{tiwari2019using} and have dynamics that develop on short time scales \cite{feng2008self}. These properties make MEMS/NEMS devices suitable for study in the field of experimental nonlinear dynamics. Examples of such studies include the demonstration of novel dynamical states such as weak chimeras and traveling waves in a ring of eight NEMS oscillators \cite{matheny2019exotic}, synchronization in optomechanical cavity oscillators \cite{colombano2019synchronization}, and frequency fluctuations in MEMS resonators \cite{pandit2021experimental}. Reviews of the study of nonlinear dynamics using micro- and nano-scale resonators are given in \cite{rhoads2008nonlinear, hajjaj2020linear}. Synchronization in coupled oscillators as a nonlinear dynamical phenomenon has also been extensively studied using theoretical and numerical analysis \cite{osipov2007synchronization, pikovsky2001universal}. 

In this work, we study the dynamics of two mechanically coupled laser-driven silicon MEMS limit cycle oscillators (LCOs) and the frequency detuning and coupling conditions in which the oscillators synchronize at a common locking frequency. Synchronization has been observed in MEMS oscillators that are optically driven and optically coupled \cite{zhang2012synchronization}, piezoelectrically driven and electronically coupled \cite{matheny2014phase} etc. Our work presents the phenomenon of synchronization in a pair of MEMS oscillators that are opto-thermally driven and mechanically coupled and extends the variety of micro-devices studied in the literature. 

In Section \ref{sec:Experiments}, the fabrication process of the MEMS devices and experimental procedures to drive and detect oscillations are discussed briefly followed by a discussion on the experimental results on the synchronization characteristics of the devices. In Section \ref{sec:Num}, the lumped-parameter mathematical model and the numerical procedures to solve the model are discussed briefly followed by a discussion on the numerical results on synchronization regions using parametric sweeps. The Arnold tongue of synchronization in the frequency detuning vs. coupling strength parameter space is charted both experimentally and numerically. The Arnold tongue is a plot in the frequency detuning vs. coupling strength parameter space with information on whether the oscillators synchronize and is used widely in the literature to characterize the synchronization properties of a system \cite{lee2014entanglement, osipov2007synchronization}. The mathematical model reveals a region of out-of-phase synchronization at low coupling strengths and low frequency detuning and is discussed in Subsection \ref{ssec:NumResultsOP}. Out–of-phase synchronization where the frequencies of the oscillators are identical and the phases are opposite, has also been recorded in other small coupled oscillator networks \cite{wu2012anti, shayak2020coexisting, vathakkattil2020limits, goldsztein2022coupled}. It is not possible to distinguish between the out-of-phase and in-phase synchronization in our experiments since only the frequency spectrum is recorded and thus we study these phenomena numerically. %Finally, in Section \ref{sec:disc}, we report results on the decrease in frequency fluctuations in coupled oscillators with an increase in coupling strength which is in agreement with similar studies performed on different devices and mathematical models \cite{chang1997phase, agrawal2013observation, antonio2012frequency}.

\section{Experiments: synchronization in coupled micromechanical LCOs} \label{sec:Experiments}
In this section, we describe the MEMS devices of interest, the fabrication techniques to build them, the experimental procedures used to analyze their dynamics, and the experimental results on the synchronization regions in the frequency detuning vs. coupling strength parameter space. We observe that the coupling strength threshold for synchronization of oscillations increases with an increase in detuning levels. At a fixed frequency detuning level, as the coupling strength is increased, we note the transition of the oscillators from a state of drift, where the frequencies of the oscillators are different, to a state of quasi-periodicity, where the oscillators differ in frequency but there is a strong amplitude modulation due to the coupling effects, to a state of synchronization where the frequencies of the oscillators are identical.
\subsection{Device design and fabrication} \label{ssec:Devices}
In this work, we study pairs of clamped-clamped micro-scale silicon beams that are coupled to each other via elastic overhangs or short bridges. The devices were fabricated on a silicon-on-insulator (SOI) $0.5''\times 0.75''$ chip with a $205$ \si{nm} thick silicon device layer, a $400$ \si{nm} thick silicon dioxide layer, and a thick silicon substrate underneath. The axis of the clamped beams are oriented along the $[100]$ crystallographic direction of the device layer.  A standard photolithography procedure was used to etch and release the devices. A positive photoresist layer was spun on the SOI chip which was then exposed in a $5\times$ g-line wafer stepper and developed. The silicon device layer was etched in an inductively coupled plasma/reactive ion etcher using \ch{C_4F_8}\ch{/SF_6} plasma. Following the plasma silicon etch, the silicon dioxide layer was etched using a $15$-minute $6:1$ buffered oxide etch to release the beams. Critical point drying was used immediately after the buffered oxide wet etch to avoid stiction effects. Optical profilometry of the micro-beams revealed that the devices are buckled outwards due to a release of axial compressive residual stresses. The amplitude-frequency response of the beams in this post-buckled state shows amplitude softening and results in a negative cubic stiffness nonlinearity in the mathematical model.

\begin{figure}
\centering
\includegraphics[width=0.35\textwidth]{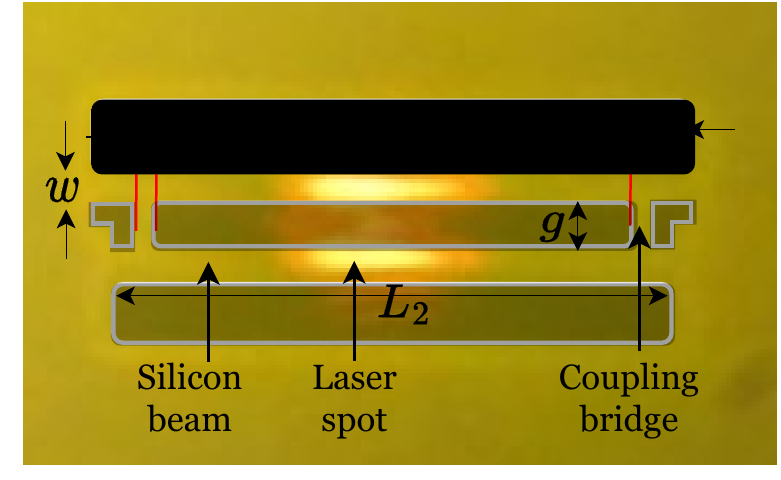}
\caption{Optical microscope image taken at $30 \times$ magnification of a sample device with the beams outlined and key dimensions labelled; $L_1 = 40$ \si{ \mu m}$, L_2 = 38$ \si{\mu m}$, w =2$ \si{\mu m}$, g = 3$ \si{\mu m}$, l = 3$ \si{\mu m}, and $v = 1$ \si{\mu m}. The out-of-plane thickness of the silicon device layer is $205$ \si{nm}. The laser spot is aimed at the center of the device to cover both beams and is used to drive and detect oscillations simultaneously.}
\label{fig:LaserSpotDimensions}
\end{figure}

Coupled beams of various lateral dimensions were fabricated on the chip. In this study, we analyzed beams of lengths $\{34, 36, 38, 40, 42\}$ \si{\mu m} and of width $w=2$ \si{\mu m}. Beams of different nominal lengths, $L$, have different baseline resonance frequencies, $f$, (which vary as $f \propto 1/L^2$ \cite{weaver1991vibration}) and limit cycle frequencies. Pairs of beams were fabricated at a lateral separation of $g=3$ \si{\mu m} as shown in Fig.~\ref{fig:LaserSpotDimensions}. The oscillators were coupled using short bridges that are beams connecting pairs of oscillators and placed symmetrically distant from the anchor points. The closer the bridges are to the center of the oscillators the stronger the coupling strength. In this study, we consider five different coupling levels namely coupling level $\{0,1,2,3,4\}$ corresponding to the distance in \si{\mu m} of the bridges from the end of the beams. In a minimally coupled case, level $0$, the bridges are absent and the coupling is mediated through the elastic overhangs resulting from the undercutting of the silicon dioxide layer during the buffered oxide etch. At the highest coupling level $4$, the coupling bridges are each at a distance of $l=4$ \si{\mu m} from the nearest anchor. The beams are driven into limit cycle oscillations using a single continuous-wave (CW) laser beam. The laser spot diameter is larger than $7$ \si{\mu m} such that it covers both the beams when the laser is directed at the center of the devices. An optical microscope image of a representative device taken at $30\times$ magnification with coupled oscillators of lengths $L_1=40$ and $L_2=38$ \si{\mu m}, and at coupling level $3$ (i.e. $l=3$ \si{\mu m}) is shown in Fig.~\ref{fig:LaserSpotDimensions}. In this figure, the device (out of focus) is outlined and the laser spot (in focus) is centered and is covering both the beams. The etched area in the device layer has been shaded and explicit coupling bridges are shown.

\begin{figure}
\centering
\includegraphics[width=0.4\textwidth]{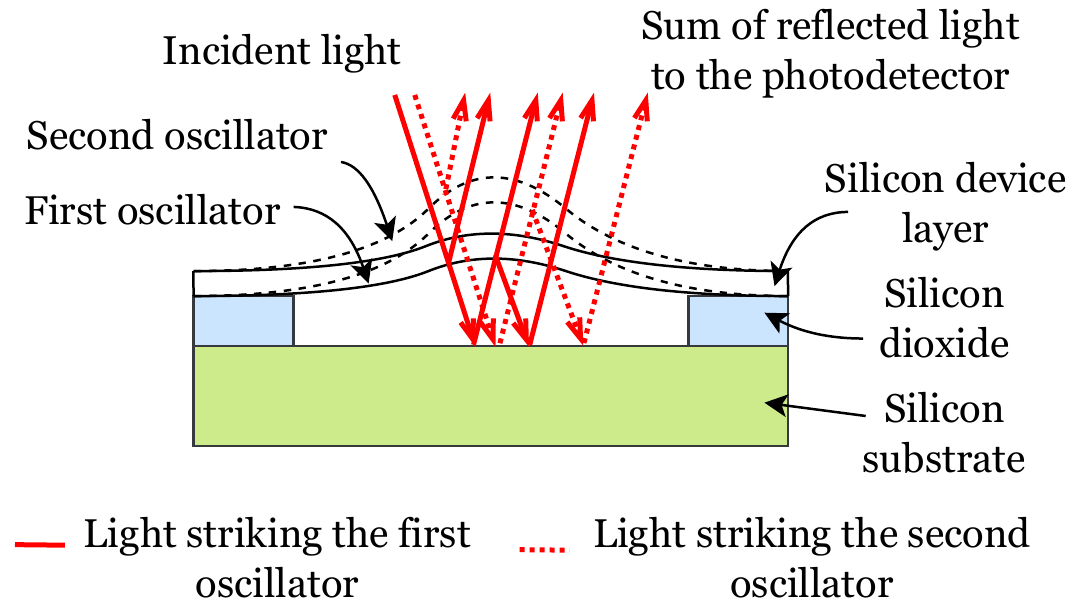}
\caption{Schematic of a pair of mechanically coupled, doubly clamped, silicon micro-oscillators on an silicon-on-insulator (SOI) chip. The device-gap-substrate forms a Fabry–Pérot interferometer. The continuous-wave (CW) laser drives the silicon beams into out-of-plane limit cycle oscillations \cite{blocher2012anchor}.}
\label{fig:FabryPerotPairSchematic}
\end{figure}

The micro-beams are driven into out-of-plane limit cycle oscillations interferometrically \cite{carr1998measurement}. The incident laser light is partly absorbed by both the beams. The part of the light that is transmitted is reflected from the silicon substrate underneath and partly reabsorbed by the beams. There are multiple rounds of reflection from the silicon substrate and absorption by the silicon beams. This sets up a Fabry-Perot interferometer in the silicon device - air gap - silicon substrate system such that the net absorbed energy and the net reflected light are both periodic functions of the gap between the beams and the substrate underneath. As the absorbed light causes thermal expansion of the devices and hence an out-of-plane motion, the absorbed light is modulated and the thermal feedback results in self-sustaining oscillations termed here as \textit{limit cycle oscillations}. The laser power should be above a threshold value for the Hopf bifurcation to occur in the system resulting in a stable limit cycle \cite{aubin2004limit}. The same scheme also is used to detect oscillations as the intensity of the net reflected light i.e. the sum of intensities of the reflected light from both the beams, is modulated at the same frequency as the limit cycle oscillations. The Fabry-Perot interference system with the device stack and laser reflections and absorptions is shown in Fig.~\ref{fig:FabryPerotPairSchematic}.

\subsection{Setup and procedures}\label{ssec:Setup}
\begin{figure}
\centering
\includegraphics[width=0.45\textwidth]{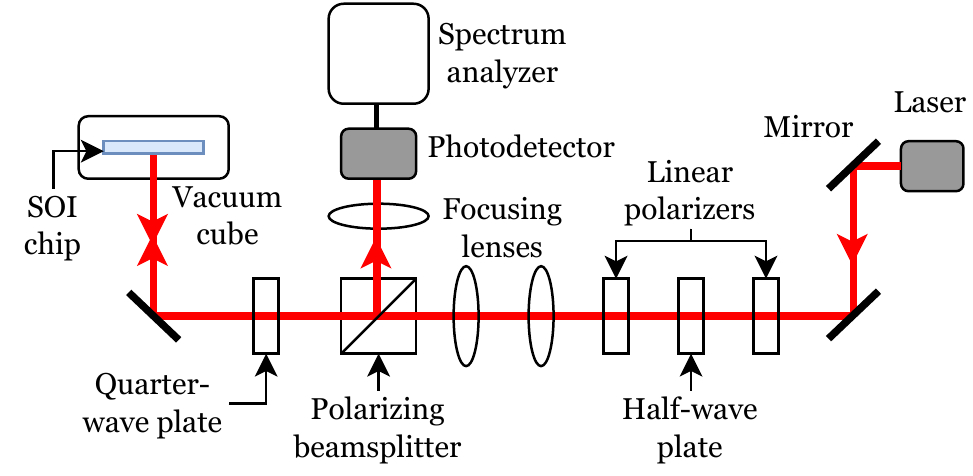}
\caption{Simplified schematic of the experimental setup to drive and detect limit cycle oscillations in the MEMS device. Laser at a fixed power and polarization is incident on the SOI chip housed in a vacuum chamber. The reflected light is directed to a photodetector and recorded on a spectrum analyzer \cite{carr1998measurement}.}
\label{fig:SimplifiedSetup}
\end{figure}

The experimental setup to drive and detect oscillations in the silicon micro-beams is shown in Fig.~\ref{fig:SimplifiedSetup} \cite{carr1998measurement}. A $20$ \si{mW} helium-neon laser source operating at $633$ \si{nm} was directed at the devices and aligned using two steering mirrors in a Z-fold configuration. A system of two linear polarizers and a half-wave plate was used to set the plane of polarization (p-polarization) and the power of the light going into the beam splitter cube. In this setup, p-polarization corresponds to the plane of polarization parallel to the optical table. The laser spot diameter was reduced using a pair of plano-convex lenses in a beam reducer configuration. A polarizing beam splitter and a quarter-wave plate constituted an optical isolator allowing circularly polarized light to enter the microscope body. The laser was focused at the center of the coupled beams using the microscope optics.

The chip was held inverted in a vacuum chamber at a high vacuum pressure of $< 10^{-6}$ \si{mBar}. The vacuum environment is needed to reduce air damping. The sealed chamber also keeps the devices free from dust, water or other contaminants. The vacuum was created by first bringing the pressure to $\approx 10^{-3}$ \si{mBar} using a cryogenic sorption pump and then further lowering the pressure to $< 10^{-6}$ \si{mBar} using an ion pump. For simplicity, the vacuum setup is not shown in Fig.~\ref{fig:SimplifiedSetup}.

The reflected light signal modulated by the out-of-plane oscillations of the micro-beams was collected in an AC-coupled high speed photodetector and recorded on a spectrum analyzer. The reflected light from the sample is the sum of intensities of the light reflected from both the micro-beams and hence the recorded spectrum contains the superposition of the limit cycle response of both the oscillators. For the three states of oscillations, the frequency spectra are qualitatively different. For drifting oscillators i.e. when non-identical oscillators are oscillating at different frequencies, the frequency spectrum of the reflected signal has two distinct peaks corresponding to the two frequencies of oscillation. For oscillators exhibiting quasi-periodic behavior, the oscillations have multiple satellite peaks which are also seen in the reflected signal. Finally, in the synchronized state, the oscillators have the same frequency and the reflected signal has a single prominent peak at the frequency of locking. The three states of oscillation and the synchronization region are discussed in Subsection \ref{ssec:ExpRes}.

\subsection{Experimental results}\label{ssec:ExpRes}
The limit cycle oscillation responses were recorded for $25$ different devices consisting of pairs of oscillators at $5$ different coupling strengths and $5$ different frequency detuning levels. In all $25$ pairs, the reference oscillator was of length $38$ \si{\mu m}. The other oscillators in the pair were of length $\{42, 40, 38, 36, 34\}$ \si{\mu m} corresponding to a minimally coupled frequency detuning percentage of $\{-18 \%, -10\%, 0\%, +10\%, +14\%\}$ respectively. All micro-beams were of width $2$ \si{\mu m}. Five coupling strengths (levels) $\{0,1,2,3,4\}$ in increasing order correspond to an increasing distance of the coupling bridges from the anchors, $l=\{0,1,2,3,4\}$ \si{\mu m}. The coupling bridges were of lateral width $v=1$ \si{\mu m}. The case of $l=0$ \si{\mu m} corresponds to minimal coupling where the coupling bridge is absent. Coupling in this case is mediated through the silicon overhang produced due to undercutting in the silicon dioxide etch step. All experiments were performed at a fixed laser power of $\approx 3.7$ \si{mW} going into the microscope port. The laser power reaching the device is $\approx 35\%$ of the power entering the microscope based on a previous study with a similar setup \cite{blocher2012optically}. Thus, in the present study, the laser power striking the device is $\approx 1.3$ \si{mW} which is higher than the threshold laser power for limit cycle oscillations. The power is also below the threshold for irregular oscillations recently demonstrated in experiments \cite{bhaskar2022bistability}. The spectra were recorded for a frequency span of $500$ \si{kHz} and a resolution bandwidth of $1$ \si{kHz}.

\begin{figure}
\centering
\includegraphics[width=0.35\textwidth]{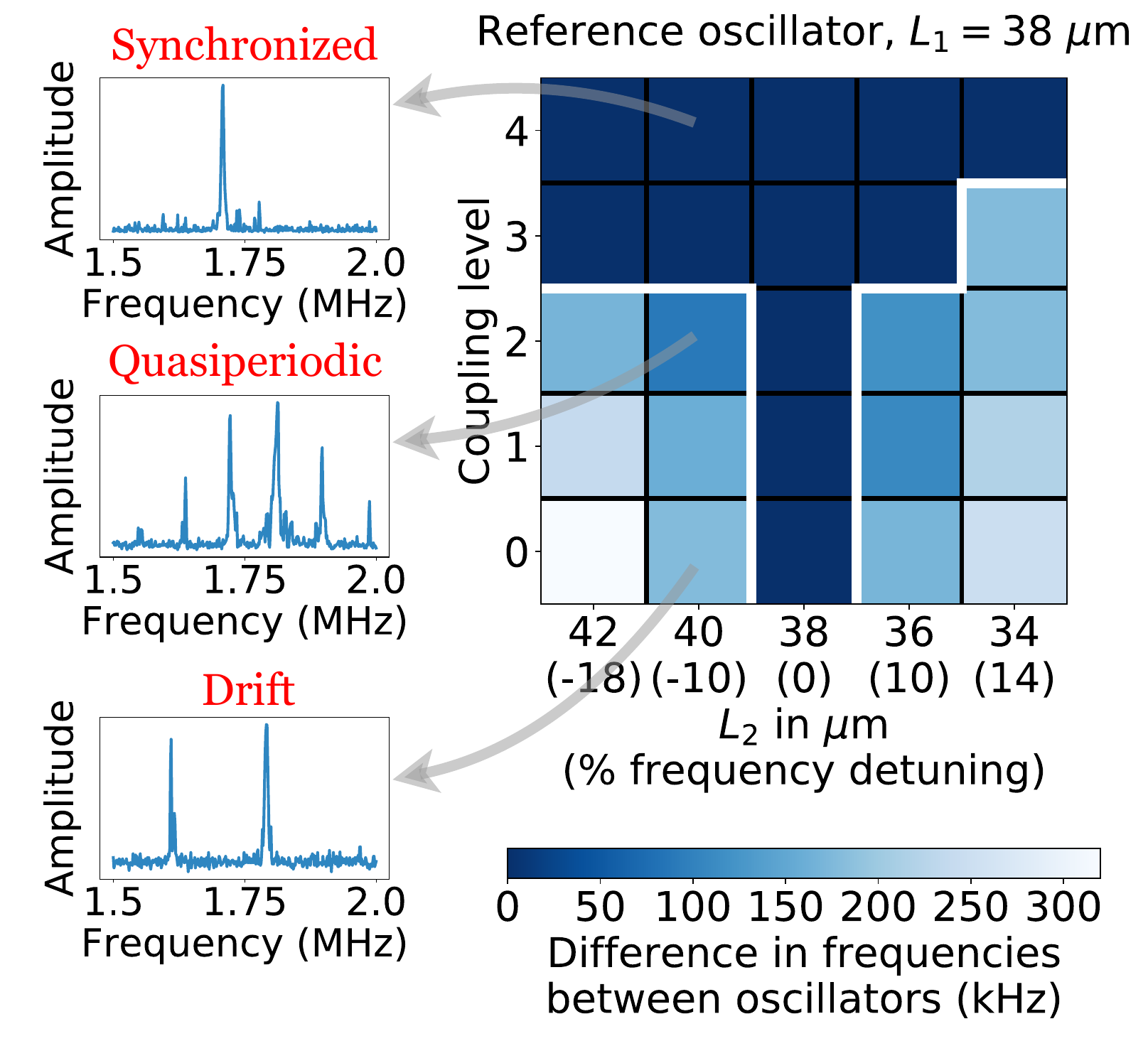}
\caption{Experimental results showing the difference in limit cycle frequencies between pairs of oscillators for various uncoupled frequency detuning and coupling strengths. The region of synchronization within the white border is termed the Arnold tongue diagram. The recorded spectra at a fixed detuning of $-10\%$ for various coupling strengths are also shown. The system moves through the drift, quasi-periodic, and synchronization states with an increase in coupling strength.}
\label{fig:ExperimentalResults}
\end{figure}

A heat map of the difference in frequencies between the two oscillators for the $25$ different devices is shown in the $5 \times 5$ grid in Fig.~\ref{fig:ExperimentalResults}. In the heat map, the difference in frequencies is visualized using a color gradation and the corresponding color bar is given. Synchronization corresponds to the region where the difference in frequencies is $0$ \si{kHz} and is bounded by the white border in the plot. Devices with a higher frequency detuning require a higher coupling strength to synchronize. The well-shaped synchronization region centered at zero detuning is similar to results for various coupled oscillator systems in the literature \cite{pikovsky2001universal} and is referred to as the Arnold tongue plot. The recorded frequency spectra corresponding to micro-beams of lengths $38$ and $40$ \si{\mu m} at coupling levels of $0, 2$ and $4$ are also given in Fig.~\ref{fig:ExperimentalResults}. The amplitudes in the frequency spectra are plotted in log-scale. At this detuning of $-10\%$ and coupling level $0$, the measured spectrum has two distinct peaks at $1.61$ and $1.79$ \si{MHz} suggesting drifting oscillators. At coupling level $2$, the measured spectrum has two prominent peaks at $1.72$ and $1.81$ \si{MHz} along with multiple satellite peaks suggesting quasi-periodic oscillations. In quasi-periodic oscillations, the amplitude of the oscillators modulate due to moderate coupling interactions. At the highest coupling level $4$, the measured spectrum collapses to a single peak at $1.71$ \si{MHz} suggesting that the oscillators are synchronized. At this coupling level the oscillators are synchronized for all five frequency detuning levels measured in this work. In a previous study \cite{bhaskar2022bistability}, we also experimentally demonstrated that the devices can show quasi-periodic and drift states at twice the input laser power. The higher laser power caused an increase in the effective frequency detuning between the oscillators allowing the system to exhibit out-of-synchronous behavior and bistability of states. We note that quasi-periodic oscillations were observed only in one out of the $25$ devices studied at this laser power. If the coupling strength could be varied continuously, then a transition from quasi-periodicity to synchronization could be observed at all frequency detuning levels.

\section{Numerical analysis:  synchronization in coupled micromechanical LCOs}\label{sec:Num}

In this section, we mathematically model, and numerically solve the nonlinear dynamics of pairs of linearly coupled opto-thermal limit cycle oscillators at various frequency detuning and coupling strengths. The third-order lumped-parameter model for the MEMS system is described and frequency spectra and phase portraits are analyzed. The phase portraits allow for distinguishing between in-phase and out-of-phase synchronization behavior and both these regimes are located in the detuning vs. coupling strength parameter space.

\subsection{Model and numerical methods}\label{ssec:ModelandMethods}
We model each limit cycle oscillator using a system of third-order ordinary differential equations \cite{zehnder2018locking}. To this, we add a linear coupling term between the oscillators to model the elastic interaction via the silicon bridges with coupling strength $\zeta$. Note that the thermal coupling between the oscillators has been ignored due to the small size of the coupling elements. The time variable, $t$, is normalized by the small amplitude period of oscillations. The out-of-plane oscillations are described by the displacement of the midpoint, $z(t)$, of the clamped-clamped beam normalized by the wavelength of the laser light. The temperature variation of the oscillators resulting in opto-thermal feedback and stable limit cycle oscillations is described by the average temperature of the beam, $T(t)$. The oscillator model is a modified Duffing equation with additional temperature feedback terms which include the temperature-dependent linear stiffness with coefficient $C$, and static displacement with coefficient $D$. The devices were kept in high vacuum, $<10^{-6}$ \si{mBar}, and experienced low viscous damping resulting in a high-Q resonator. The cubic stiffness $\beta$ was chosen to model strong amplitude-softening nonlinearity while maintaining stable oscillations for the corresponding linear stiffness $\kappa$. The temperature variation of the beams was modeled using Newton’s law of cooling and a trigonometric approximation of the laser absorption curve for a $205$ \si{nm} thick silicon layer interacting with a $633$ \si{nm} laser and varying gap thickness. The laser power is fixed at $P_{\text{laser}}=1.5$ \si{mW} for all the numerical analysis in this work which is comparable to the $\approx 1.3$ \si{mW} laser power used in the experiments. The model parameters are listed in Table \ref{tab:tab1} \cite{blocher2012optically}.

\begin{align}
&\ddot{z}_1 + \frac{\dot{z}_1}{Q} + \kappa_1(1+CT_1)z_1 + \beta z_1^3 + \zeta(z_1-z_2) =DT_1, \label{eq: displ1} \\
&\dot{T}_1 = -BT_1 + HP_{\text{laser}}\left(\alpha + \gamma\sin^2(2\pi(z_1-\bar{z}))\right), \label{eq: temp1}\\
&\ddot{z}_2 + \frac{\dot{z}_2}{Q} + \kappa_2(1+CT_2)z_2 + \beta z_2^3 + \zeta(z_2-z_1) =DT_2, \label{eq: displ2} \\
&\dot{T}_2 = -BT_2 + HP_{\text{laser}}\left(\alpha + \gamma\sin^2(2\pi(z_2-\bar{z}))\right). \label{eq: temp2}
\end{align}
\begin{table*}
\centering
\begin{tabular}{llll}
\hline\noalign{\smallskip}
Parameter & Symbol & Value & Units  \\
\noalign{\smallskip}\hline\noalign{\smallskip}
Quality factor & $Q$ & $1240$ & [AU] \\
Cubic nonlinearity & $\beta$ & $-10$ & [AU]\\
Coefficient of static displacement per unit temperature & $D$ & $2.84 \times 10^{-3}$ & \si{1/K} \\
Coefficient of linear stiffness per unit temperature & $C$ & $0.04$ & \si{1/K} \\
Coefficient of heat transfer & $B$ & $0.112$ & [AU] \\
Minimum absorption & $\alpha$ & $0.035$ & [AU] \\
Contrast in absorption & $\gamma$ & $0.011$ & [AU]  \\
Minimum of absorption w.r.t device equilibrium & $\bar{z}$ & $0.18$ & [AU] \\
Coefficient of thermal absorption  & $H$ & $6780$ & \si{K/W} \\
Laser power  & $P_{\text{laser}}$ & $1.5$ & \si{mW} \\
%$P_{\text{laser}}$ & $2 \times 10^{-3}$ & W \\
\noalign{\smallskip}\hline{\smallskip}
\end{tabular}
\caption{Model parameters used in Eqs.~(\ref{eq: displ1})-(\ref{eq: temp2}). [AU] stands for arbitrary units.}
\label{tab:tab1}
\end{table*}
The mathematical model was numerically integrated in Python using lsoda and a parallelized code for the parameter sweeps. The steady-state time series $z(t)$ for each limit cycle oscillator was used to compute the Fast Fourier Transform to a frequency resolution of $1.5 \times 10^{-6}$ [AU]. In the drift state, the oscillators have non-identical frequencies which are represented by two prominent peaks in the spectrum. In the quasi-periodic state, the oscillators have non-identical frequencies and the spectrum for each oscillator has multiple satellite peaks. The satellite peaks are a result of the moderate coupling interaction between the two oscillators. In the synchronized state, both the oscillators have a single prominent peak at the same frequency. The steady-state time series $z_1(t)$ and $z_2(t)$ were also used to plot the phase portrait $z_1(t)$ vs. $z_2(t)$. In the drift state, the data points fill the entire phase portrait. In the quasi-periodic state, the data points fill out a restricted region in the phase portrait. Qualitatively, the drift and quasi-periodic states of oscillations differ in the amplitude modulation. In the presence of minimal coupling, the oscillators have nearly constant oscillation amplitudes. In the presence of moderate coupling, there is significant amplitude modulation in the oscillators in the quasi-periodic state. In the synchronized state, the phase portrait is a straight line passing through the origin since the oscillator frequencies are identical. For in-phase synchronization, the slope of the phase portrait is positive, and for out-of-phase synchronization, the slope is negative. Note that in-phase and out-of-phase synchronization can be distinguished using the phase portrait but not the Fourier spectrum. Since the spectrum is the only data that we record in the experiments, we are able to report in experiments synchronization of the oscillators but not whether the oscillators are synchronized in-phase or out-of-phase. All numerical results in this section are presented in the form of parametric sweeps \cite{towns2014xsede} in the frequency detuning, determined by the linear stiffness $\kappa_2$, and the coupling strength between the oscillators $\zeta$. For each pair of parameters $(\kappa_2, \zeta)$, the model is numerically integrated for $25$ randomized initial conditions and the probability of synchronization (in-phase or out-of-phase) is plotted in the form of a heat map.
\subsection{Numerical results: in-phase synchronization}\label{ssec:NumResultsIP}
\begin{figure}
\centering
\includegraphics[width=0.5\textwidth]{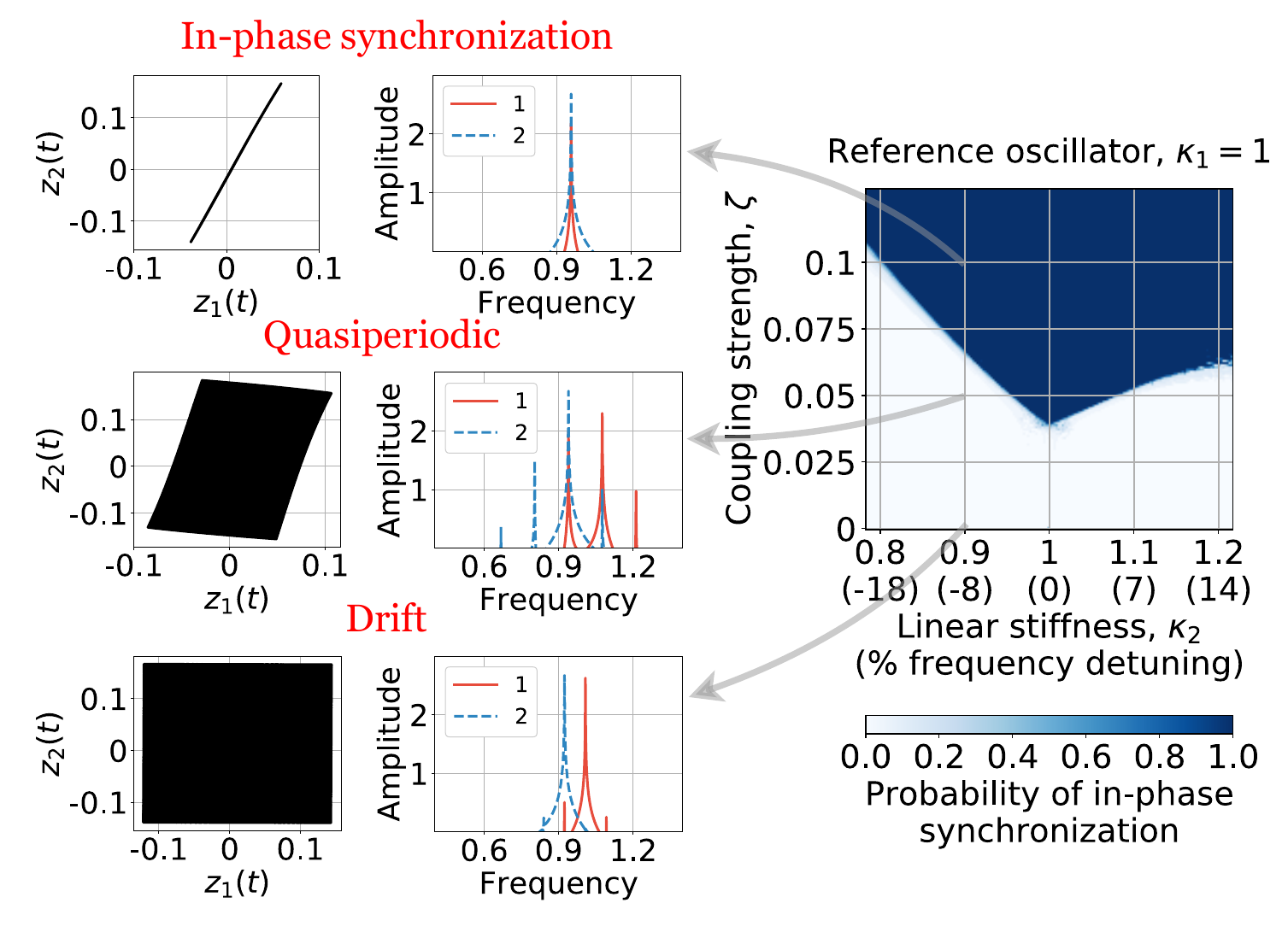}
\caption{Numerical results showing the probability of \textit{in-phase synchronization} for various uncoupled frequency detuning and coupling strengths. The phase portraits and spectra at a fixed detuning of $-8\%$ for various coupling strengths are also shown. The system moves through the drift, quasi-periodic, and synchronization states with increasing coupling strengths at a fixed detuning level.}
\label{fig:IPProbab}
\end{figure}

The probability distribution for in-phase synchronization in the frequency detuning vs. coupling strength space is given in Fig.~\ref{fig:IPProbab}. The linear stiffness term for the reference oscillator is fixed at $\kappa_1=1$ and the uncoupled frequency detuning between the oscillators were varied between $\kappa_2 \approx 0.8$ and $\kappa_2 \approx 1.2$. This range of linear stiffness values corresponds to numerical uncoupled frequency detuning percentages from $-18 \%$ to $+14 \%$ which match the range in the experiments. The coupling strength between the oscillators varied between $\zeta=0$ to $\zeta \approx 0.12$. All the numerical calculations were performed at a fixed laser power of $P_{\text{laser}}=1.5$ \si{mW}. To plot the probability of in-phase synchronization, the initial conditions for the numerical calculations were randomized in the interval $z_1(0) \in [-0.1, 0.1)$ and $z_2(0) \in [-0.1, 0.1)$. All other initial conditions were fixed at $\dot{z}_1(0)=\dot{z}_2(0)=T_1(0)=T_2(0)=0$. The probability distribution for synchronization shows a well-shaped region of synchronization where a higher coupling strength is required for synchronization at a higher frequency detuning. This region is the Arnold tongue corresponding to the numerical calculations. The Fourier spectra and the phase portrait are shown for a fixed frequency detuning of $\kappa_2=0.9$ ($-10\%$) and three coupling strengths i.e. low coupling $\zeta=5 \times 10^{-4}$, moderate coupling $\zeta=0.05$, and strong coupling $\zeta=0.1$. The amplitudes in the Fourier spectra are plotted using Arbitrary Units [AU] and in log-scale. The oscillators are drifting at low coupling, show quasi-periodic behavior at moderate coupling, and exhibit synchronization at high coupling. These numerical results can be compared to the experimental results in Fig.~\ref{fig:ExperimentalResults}. The plot in Fig.~\ref{fig:ExperimentalResults} has a heat map of the difference in frequencies whereas the plot in Fig.~\ref{fig:IPProbab} has a heat map of the probability of in-phase synchronization. In the numerical simulations, we chose to plot the probability distribution instead of the difference in frequencies since the probability map also reveals the sensitive dependence of the system on initial conditions. Both the plots show a qualitatively similar boundary for the synchronization region. It is to be noted that in-phase synchronization is not observed in the numerical calculations at low coupling strengths ($\zeta<0.025$) and even for low detuning (and identical oscillators, $\kappa_2=1$). At low coupling strengths and low detuning the oscillators tend to synchronize out-of-phase.   

\subsection{Numerical results: out-of-phase synchronization}\label{ssec:NumResultsOP}
\begin{figure}
\centering
\includegraphics[width=0.5\textwidth]{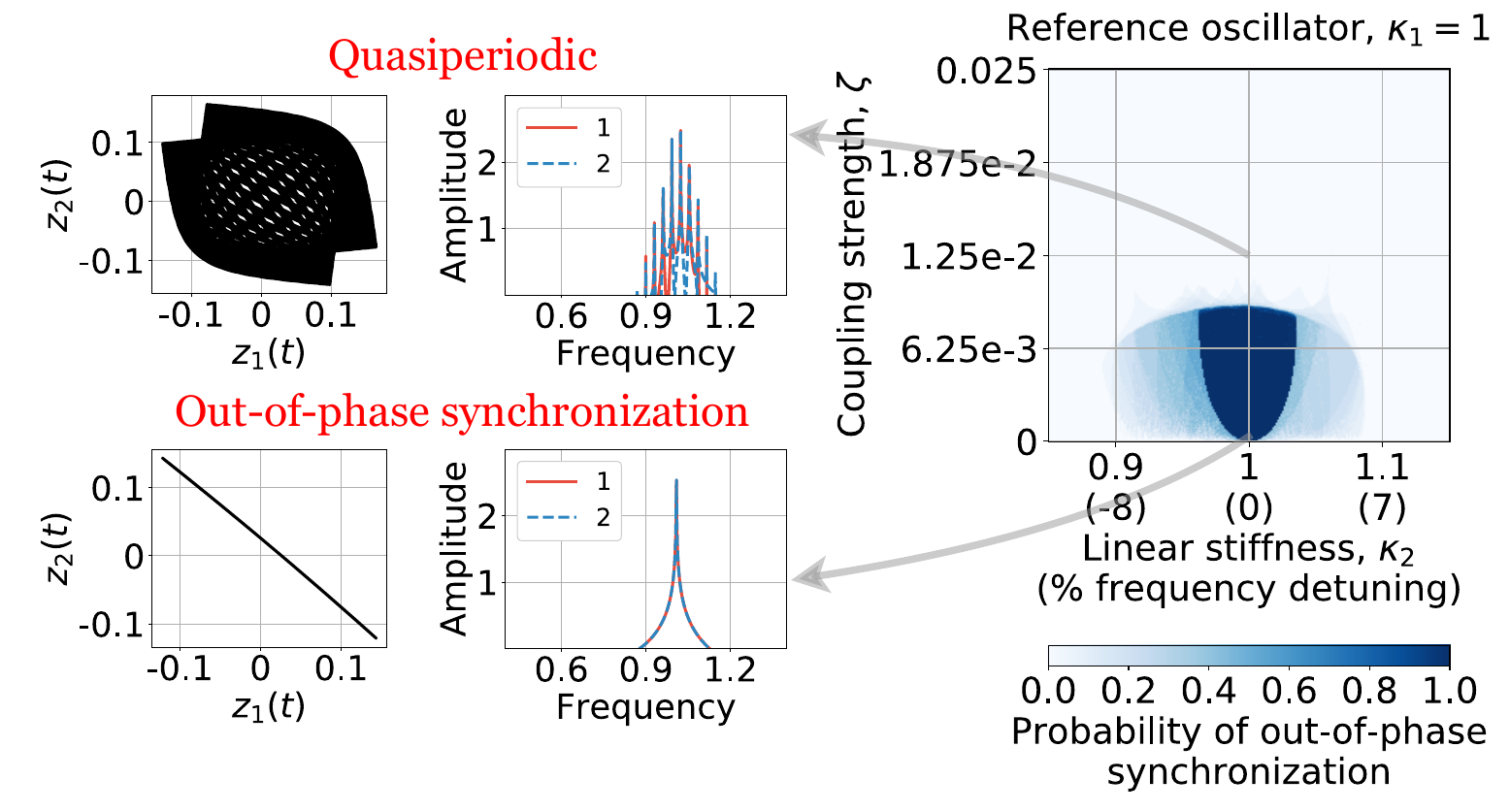}
\caption{Numerical results showing the probability of \textit{out-of-phase synchronization} for various uncoupled frequency detuning and coupling strengths. The phase portrait and spectra for identical oscillators at two coupling strengths are also shown. The system moves from out-of-phase synchronization to the quasi-periodic state with increasing coupling strengths. The light blue region outside the boundaries of certain synchrony correspond to a state of bistability, where the oscillators may or may not synchronize out-of-phase depending on the initial conditions.}
\label{fig:OPProbab}
\end{figure}

The probability distribution for out-of-phase synchronization in the frequency detuning vs. coupling strength space is given in Fig.~\ref{fig:OPProbab}. The linear stiffness for the reference oscillator was again fixed at $\kappa_1=1$ and that of the second oscillator varied in a smaller range; between $\kappa_2 \approx 0.9$ and $\kappa_2 \approx 1.1$. The coupling strength between the oscillators varied in a smaller range; between $\zeta=0$ to $\zeta = 0.025$. Again, all the numerical calculations were performed at a fixed laser power of $P_{\text{laser}}=1.5$ \si{mW}. The initial conditions for the numerical calculations were randomized in the interval $z_1(0) \in [-0.1, 0.1)$ and $z_2(0) \in [-0.1, 0.1)$. All other initial conditions were fixed at $\dot{z}_1(0)=\dot{z}_2(0)=T_1(0)=T_2(0)=0$. We observe a symmetric region of out-of-phase synchronization only for low detuning and low coupling strengths centered at zero detuning. Outside the region of certain out-of-phase synchronization, there is a bistable region where the oscillators could either exhibit out-of-phase synchronization or drift/quasi-periodic behavior and the region is marked in light blue. Such regions characterized by fuzzy boundaries have been numerically observed in prior work \cite{zehnder2018locking, bhaskar2021synchronization}. For the case of identical oscillators, as the coupling strength is increased, the oscillators go from out-of-phase synchronization to quasi-periodic behavior as shown in the Fourier spectra and phase portraits in Fig.~\ref{fig:OPProbab}. From Fig.~\ref{fig:IPProbab}, we note that further increase of the coupling would lead to in-phase synchronization of the oscillators.

\section{Discussion}\label{sec:disc}
In Sections \ref{sec:Experiments} and \ref{sec:Num}, we showed using Arnold tongue diagrams that the oscillators synchronize to a common frequency when the coupling strength is sufficiently high. The route to synchronization from the drift state to the quasi-periodic state to the synchronized state was observed in both the experiments as well as numerical analysis. The existence of out-of-phase synchronization was noteworthy and is consistent with existing literature. Out-of-phase synchronization is typically seen in networks with less than $20$ elements, where such synchronization is defined as two clusters of oscillators self-synchronized but opposite in phase with respect to each other \cite{vathakkattil2020limits}. Additionally, out-of-phase synchronization is seen at low coupling strengths. The transition from out-of-phase to in-phase synchronization with an increase in coupling strength has been seen in systems such as mechanical metronomes coupled via Coulomb friction \cite{goldsztein2022coupled} and coupled chemical oscillators \cite{awal2019smallest}.

\section{Conclusions}\label{sec:conc}
In this work, we studied the nonlinear dynamical phenomena exhibited by mechanically coupled pairs of opto-thermal limit cycle oscillators at various frequency detuning and coupling strengths. The synchronization characteristics of the devices driven with the novel opto-thermal actuation technique and a mechanically coupled using a unique design distinguish our work from previous studies on the dynamics of MEMS devices. We focused on the phenomenon of synchronization, recording it experimentally and numerically using Arnold tongue diagrams. Three qualitatively different states of oscillations; drift state, quasi-periodic state, and synchronized state were observed. The numerical results also revealed distinct regions of in-phase synchronization at moderate coupling strengths and out-of-phase synchronization at relatively low coupling strengths. 

\section*{Acknowledgments}
The authors would like to thank the anonymous reviewers for their suggestions and feedback.

\bibliographystyle{IEEEtran}

% \newpage
 
\vspace{11pt}

\begin{IEEEbiography}[{\includegraphics[width=1in,height=1.25in,clip,keepaspectratio]{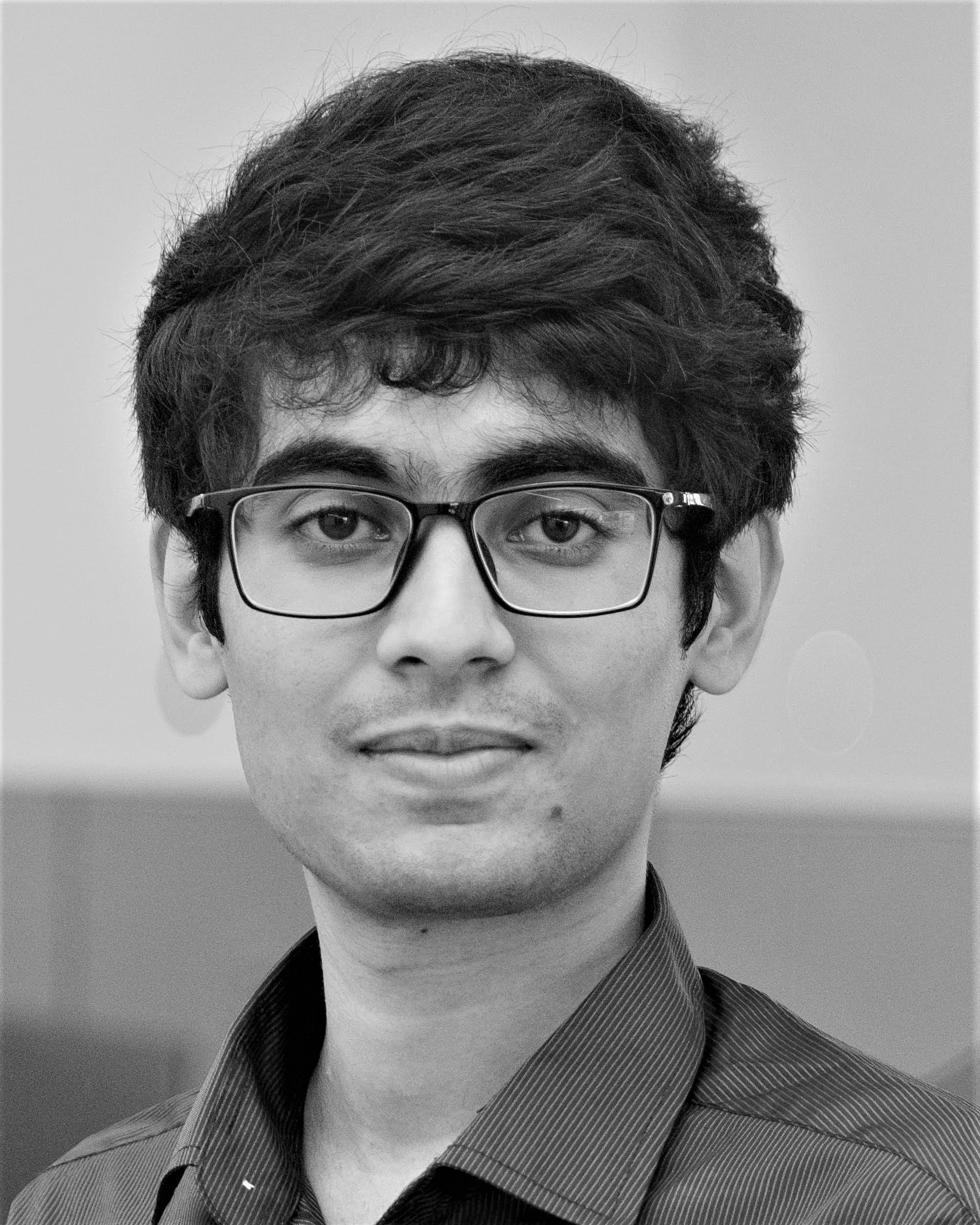}}]{Aditya Bhaskar} was born in Mumbai, India.
He received the B.Tech. and M.Tech. degrees from the Department of Mechanical Engineering, Indian Institute of Technology Madras, Chennai, India, in 2017 and the Ph.D. degree from the Sibley School
of Mechanical and Aerospace Engineering, Cornell University, Ithaca, NY, USA, in 2022. His research interests include nonlinear dynamics of MEMS oscillators, distributed averaging algorithms, and fracture propagation in heterogeneous materials.   
\end{IEEEbiography}

\begin{IEEEbiography}[{\includegraphics[width=1in,height=1.25in,clip,keepaspectratio]{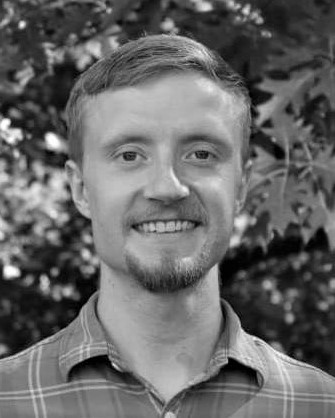}}]{Mark Walth} was born in Colorado Springs, CO, USA. He received the B.A. degree in Mathematics from Reed College, Portland, OR, USA, in 2014 and the M.A.T. degree in Mathematics Education from American University, Washington, DC, USA, in 2015. He is currently pursuing the Ph.D. degree with the Department of Mathematics, Cornell University, Ithaca, NY, USA, working on nonlinear dynamics. Prior to that, he was a middle school mathematics teacher in Washington, DC, USA, as a member of the Math for America Fellowship Program.   
\end{IEEEbiography}

\begin{IEEEbiography}[{\includegraphics[width=1in,height=1.25in,clip,keepaspectratio]{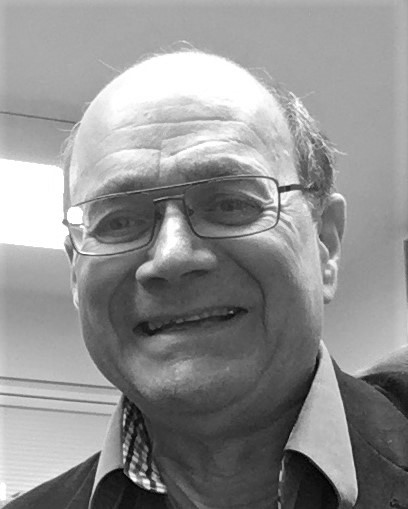}}]{Richard H. Rand} received the B.S. degree from The Cooper Union, New York, NY, USA, in 1964 and the M.S. and Sc.D. degrees from Columbia University, New York, NY, USA, in 1965 and 1967, respectively, all in Civil Engineering. He has been a Professor with Cornell University, Ithaca, NY, USA, since 1967 and is currently with both the Department of Mathematics and the Sibley School of Mechanical and Aerospace Engineering. His recent research works involve using perturbation methods and bifurcation theory to obtain approximate solutions to differential equations arising from nonlinear dynamics problems in engineering and biology.
\end{IEEEbiography}

\begin{IEEEbiography}[{\includegraphics[width=1in,height=1.25in,clip,keepaspectratio]{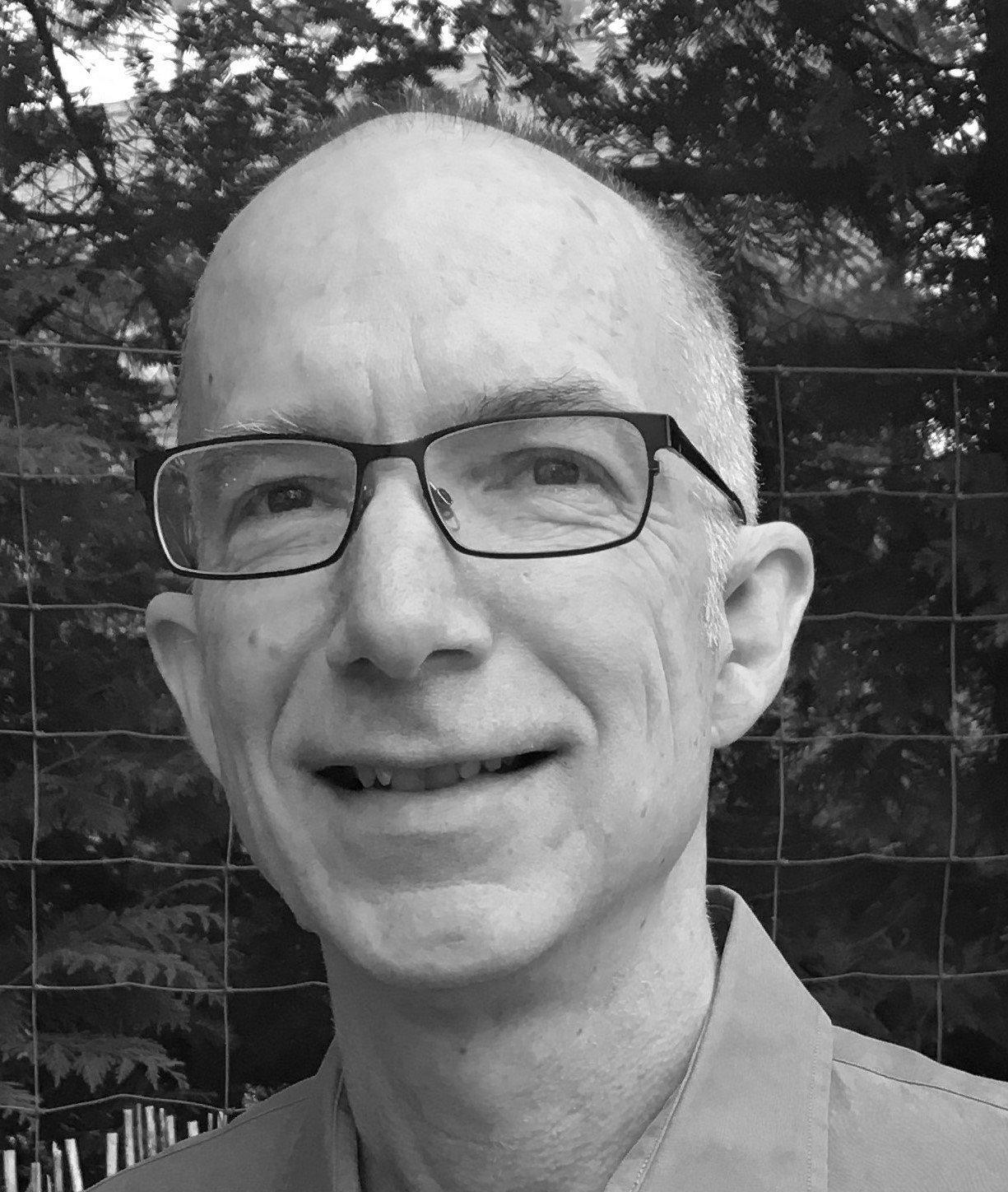}}]{Alan T. Zehnder} received the B.S. degree from the University of California at Berkeley, USA, and the Ph.D. degree from the California Institute of Technology, Pasadena, USA, both in Mechanical Engineering. He is a Professor with the Sibley School of Mechanical and Aerospace Engineering, Cornell University, Ithaca, NY, USA, where he serves as the Associate Dean for Undergraduate Programs. His current research interests include the nonlinear dynamics of nanomechanical oscillators and the deformation and fracture of hydrogels. He is a Fellow of the American Society of Mechanical Engineers (ASME) and the Society for Experimental Mechanics (SEM). He is a recipient of the S. Nemat Nassar Award from the SEM. He is the Editor-in-Chief of \textit{Experimental Mechanics}.
\end{IEEEbiography}

\vfill

\end{document}